\documentclass[useAMS,usenatbib]{mnras}
\usepackage{graphicx,amssymb,xcolor,ulem}
\usepackage{bm} 
\def\mnras{MNRAS}
\def\apjl{ApJ}
\def\apj{ApJ}
\def\apjs{ApJS}
\def\aaps{A\&AS}
\def\aap{A\&A}
\def\eb{$\epsilon_{\rm B}$}
\def\ee{$\epsilon_{\rm e}$}
\def\etag{$\eta_\gamma$}
\def\gray{$\gamma$-ray}
\def\tee{temporally extended emission}
\def\g0{$\Gamma_0$}
\def\etag{$\eta_\gamma$}

\def\ek{$E_{\rm k}$}
\def\eiso{$E_{\rm \gamma,iso}$}

\def\t90{$T_{90}$}
\def\nuc{$\nu_{\rm c}$}
\def\num{$\nu_{\rm m}$}
\def\slat{$S^{\rm aft}_{\rm [0.1-10]}$}

\def\spr{$S_{\rm \gamma,iso}$}

\title[Upper limits on GRB's bulk Lorentz factors]
  {Constraints on the bulk Lorentz factor of Gamma-Ray Burst jets from {\it Fermi}/LAT upper limits}
\author[L. Nava et al.]
  {L.~Nava$^{1,4,5}$\thanks{lara.nava@ts.infn.it},
  R.~Desiante$^{2,3}$,
  F.~Longo$^{4,5}$,
  A.~Celotti$^{4,6,7}$,
  N.~Omodei$^8$,
  G.~Vianello$^8$, \newauthor 
  E.~Bissaldi$^{9,10}$,
  T.~Piran$^1$\\
 \\
  $^1$Racah Institute of Physics, The Hebrew University of Jerusalem, Jerusalem, 91904, Israel\\
  $^2$INFN - Sezione di Torino, I-10125 Torino, Italy\\
  $^3$Universit\`a di Udine, I-33100 Udine, Italy\\
  $^4$INFN - Sezione di Trieste, via Valerio 2, I-34127 Trieste, Italy\\
  $^5$Dipartimento di Fisica, Universit\`a di Trieste, I-34127 Trieste, Italy\\
  $^6$SISSA, via Bonomea 265, I-34136 Trieste, Italy\\
  $^7$INAF -Osservatorio Astronomico di Brera, via Bianchi 46, I-23807 Merate, Italy\\
  $^8$W.W. Hansen Experimental Physics Laboratory, Kavli Institute for Particle Astrophysics and Cosmology, Department of Physics\\ and SLAC National Accelerator Laboratory, Stanford University, Stanford, CA 94305, USA\\
  $^9$Dipartimento Interateneo di Fisica, Politecnico di Bari, Via E.Orabona 4, 70125 Bari, Italy\\
  $^{10}$INFN - Sezione di Bari,Via E.Orabona 4, I-70125 Bari, Italy
}
\date{Released}

\pagerange{\pageref{firstpage}--\pageref{lastpage}} \pubyear{2012}

\begin{document}
\voffset -1truecm 

\label{firstpage}

\maketitle

\begin{abstract}
It is largely recognized that Gamma-Ray Burst (GRB) jets involve ultra-relativistic
motion. However, the value of the Lorentz factor $\Gamma_0$ is still not clear and  only lower limits are known for most bursts. 
We  suggest here a new method to obtain upper limits on $\Gamma_0$. 
The early high-energy synchrotron afterglow flux depends strongly on $\Gamma_0$. Upper limits on GeV emission  therefore provide  uppers limit on $\Gamma_0$. Applying this method to 190 {\it Fermi} GRBs  that have not been detected by the {\it Fermi}-LAT we place upper limits on the 
high-energy afterglow flux, and in turn on $\Gamma_0$.  For  bursts at a typical redshift
$z=2$, we find values of the order of 200 (and above) for a homogeneous density medium, and in the range 100-400 for a wind-like medium. 
These upper limits are consistent with (and are very close to) lower limits and direct estimates inferred using other methods, suggesting that 
the typical Lorentz factors of GRB jets are of order a few hundred. 
\end{abstract}


\section{Introduction}\label{sect:introduction}
Gamma-Ray Burst (GRB) jets move at relativistic velocities with Lorentz factors \g0\ much in excess of unity. 
The properties of the emission (such as timescales and typical frequencies) measured in the observer frame appear then very different from the intrinsic ones in the comoving frame of the fluid.
Only by estimating \g0\ it is possible to infer the intrinsic properties of the emitting region.
Unfortunately, it is difficult to place significant constraints on \g0\ from observations.
As a consequence, a lot of useful information (such as the location of the dissipation region, the ejecta mass, the typical frequencies of the emitted photons), fundamental for discriminating among different theoretical scenarios, suffer from large uncertainties. 
Improving the estimates of the Lorentz factor is then essential for understanding the nature of the central engine and outflow, the conditions at the emitting region, and the nature of the radiation process.

It was early realised that an ultra-relativistic motion is needed in order to avoid the so-called compactness problem and explain detections of \gray\ photons on short variability timescales  \citep{ruderman75,krolik91,fenimore93,piran95,baring97}. The highest photon energy detected during the prompt emission can then be used to compute the minimum value of \g0\ required to avoid $\gamma$-$\gamma$ opacity within the emitting region. 
Using this method, lower limits in the range 100-400 have been derived by \cite{lithwick01} for a sample of 13 BATSE bursts.
Much larger lower limits (in the range $900-1200$) have been derived for GRBs detected by the {\it Fermi}-LAT  \citep{080916CLAT,090902B,090510LAT}, due to the extension of the accessible range to GeV energies.
These large lower limits pose severe constraints on the baryon load of the ejecta, favouring Poynting flux dominated jets.
However,  the formula used to derive these extreme values has been questioned by \cite{hascoet12b}, who proposed a more detailed calculation of $\gamma$-$\gamma$ opacity and suggested that the simpler formula overestimates \g0\ by a factor of 2-3.
Moreover, \citet{zou11} pointed out that these limits rely on the one-zone model, where GeV and sub-MeV photons are emitted from the same region and are produced by internal shocks. The long-lasting nature of the GeV emission suggests a different origin and dissipation radius for the high-energy component.
A two-zone model (where the collisions between the GeV and MeV photons occur at larger radii than the prompt emission radius) implies much weaker constraints (about one fifth to one half of the one-zone values).

Another widely used technique for estimating \g0\ is based on the onset of the afterglow emission \citep{sari99}.
A few efforts have been made to collect samples of GRBs displaying a peak in their early time optical lightcurve and derive the value of \g0, assuming that the peak time marks the outflow deceleration time. \cite{liang10,liang15} derived values in the range 90-600.
Smaller values, between 30-300 (and between 20 and 200 for a wind-like density medium) were instead inferred by \cite{ghirlanda12}. 
When the onset is not observed, (i.e., observations start when the flux is already decaying) an upper limit can be placed on the deceleration time, and then a lower limit on the value of \g0. The lower limits derived using this method are in the range 40-300 \citep{hascoet14}.

\cite{zou10} suggested that flux limits on the early afterglow can also be used to constrain \g0.
For large \g0, indeed, the afterglow emission starts at an earlier time and has a higher peak luminosity. A lack of detection can then be translated into an upper limit on the brightness of the afterglow, and then on the value of \g0.
They considered early X-ray observations in a sample of 16 GRBs and derived upper limits on \g0\ of several hundreds.
We suggest that a similar method can be applied also to high-energy (GeV) observations.
In the standard afterglow model, the early time afterglow emission is expected to extend up to GeV energies. A lack of GeV emission can then be translated into an upper limit on \g0.

In this paper, we propose to exploit LAT flux upper limits derived on timescales longer than the prompt duration to place limits on the brightness of the synchrotron afterglow component, and in turn on \g0. 
We have already applied this method to a sample of 28 GRBs observed by AGILE \citep{longo12}, deriving values between 100 and a few thousands (for a typical redshift $z=2$).
The LAT allows us to place more stringent constraints thanks to its higher sensitivity, and to significantly increase the sample of GRBs to which this analysis can be applied (190 events).

The paper is organized as follows. In \S \ref{sec:model}, we calculate the synchrotron afterglow flux in the range 0.1-10~GeV, and provide equations that can be used to place upper limits on \g0\ from the upper limits on the LAT flux.
In \S \ref{sec:UL} we consider a sample of 190 GRBs with no LAT detection, and compute the upper limits on \g0.
Some implications for GRBs detected by the LAT are discussed in \S \ref{sec:detected}. 
Conclusions are summarised in \S \ref{sec:conclusions}.

\section{Expected High-Energy afterglow emission}\label{sec:model}
In fast cooling regime, the bolometric afterglow luminosity from the forward external shock is proportional to the rate at which the energy is dissipated at the shock $dm\Gamma^2/dt$ (where $m$ is the total mass of the external medium collected up to the time $t$) and to the fraction \ee\ of this energy gained by the accelerated electrons. 
Since we are interested in early time afterglow evolution, we assume that $\Gamma$ is larger than $1/\theta_{\rm jet}$ (where $\theta_{\rm jet}$ is the jet opening angle) and express energetics and luminosities in terms of their isotropic equivalent values.
Using $dr\propto\Gamma^2 dt$ and introducing a generic density radial profile $n=n_0 r^{-s}$, the bolometric luminosity is \citep{sari97}:
\begin{equation}
L^{\rm aft}_{\rm bol}\propto \epsilon_e t^{2-s} n_0 \Gamma^{8-2s}.
\end{equation} 
Two regimes can be identified:
\begin{itemize}
\item A coasting phase ($\Gamma=\Gamma_0$): the luminosity has a strong dependence on the value of \g0, and is proportional to $n_0$ (we consider here and elsewhere in this work that \ee\ has more or less the same value for all GRBs, see below for a discussion). 
In a constant density medium $(s=0)$ the luminosity rises as $t^2$, while it is constant for a wind-like medium ($s=2$);

\item A deceleration phase: $\Gamma$ decreases according to $\Gamma^2\propto E_{\rm k}/m(r)$ \citep{blandford76}, where \ek\ is the blastwave energy (we are assuming an adiabatic evolution, i.e. \ek=constant), and $m(r)$ is the total mass collected up to the radius $r$. 
Regardless of the radial density profile, the luminosity decreases with time as $L^{\rm aft}_{\rm bol}\propto \epsilon_{\rm e} E_{\rm k}t^{-1}$. 
Since \ek\ is related to the prompt radiated energy \eiso\ through the prompt efficiency $\eta_\gamma$ (\ek=\eiso$[1-\eta_\gamma]/\eta_\gamma)$, we can write
$L^{\rm aft}_{\rm bol}\propto\epsilon_{\rm e} E_{\rm \gamma,iso} (1-\eta_\gamma)/\eta_\gamma t^{-1}$. 
\end{itemize}

The energies we are interested in ($>0.1\,$GeV) are most likely larger than the cooling and synchrotron characteristic frequencies \nuc\ and \num.
Electrons radiating at such energies are rapidly cooling, and the equations describing the luminosity of the emitted radiation are similar to equations governing the bolometric luminosity, with minor corrections to the exponents and with the introduction of a weak dependence on the fraction of energy \eb\ in the amplified magnetic field.
In particular, during the deceleration \citep{sari98}:\newline
$L^{\rm aft}_{\rm [0.1-10]}=k\,\epsilon_{\rm e}^{p-1}\epsilon_B^{\frac{p-2}{4}} [E_{\rm \gamma,iso} (1-\eta_\gamma)/\eta_\gamma]^{\frac{p+2}{4}}  t^{-\frac{3p-2}{4}}$, where the numerical factor $k$ depends only on $p$ (the power-law index of the electron injection spectrum, $N_{\rm inj}(\gamma)\propto \gamma_e^{-p}$) and varies less than a factor 1.5 for $p$ in the range $2.1-2.8$. 

This latter equation implies that, during the deceleration, the ratio between the high-energy afterglow luminosity, at a fixed rest frame time, and the prompt energy \eiso\ depends only on two parameters, \ee\ and \etag\ \citep{kumar00,freedman01}. \citet{nava14} found that for LAT GRBs with temporally extended emission, the value of this ratio is narrowly clustered, implying that the product $\epsilon_{\rm e}^{p-1} [(1-\eta_\gamma)/\eta_\gamma]^{\frac{p+2}{4}}$ has more or less the same value in different GRBs and does not introduce a significant scatter (see \citealt{nava14} for a more detailed discussion). Hereafter, we will assume that both \ee\ and \etag\ do not vary by a significant amount, but we explicitly write how our estimates depends on these two parameters, so that the effects of a different assumption can be easily computed.

\begin{figure}
\vskip -1.3truecm
\hskip -2.9truecm
\includegraphics[scale=0.37]{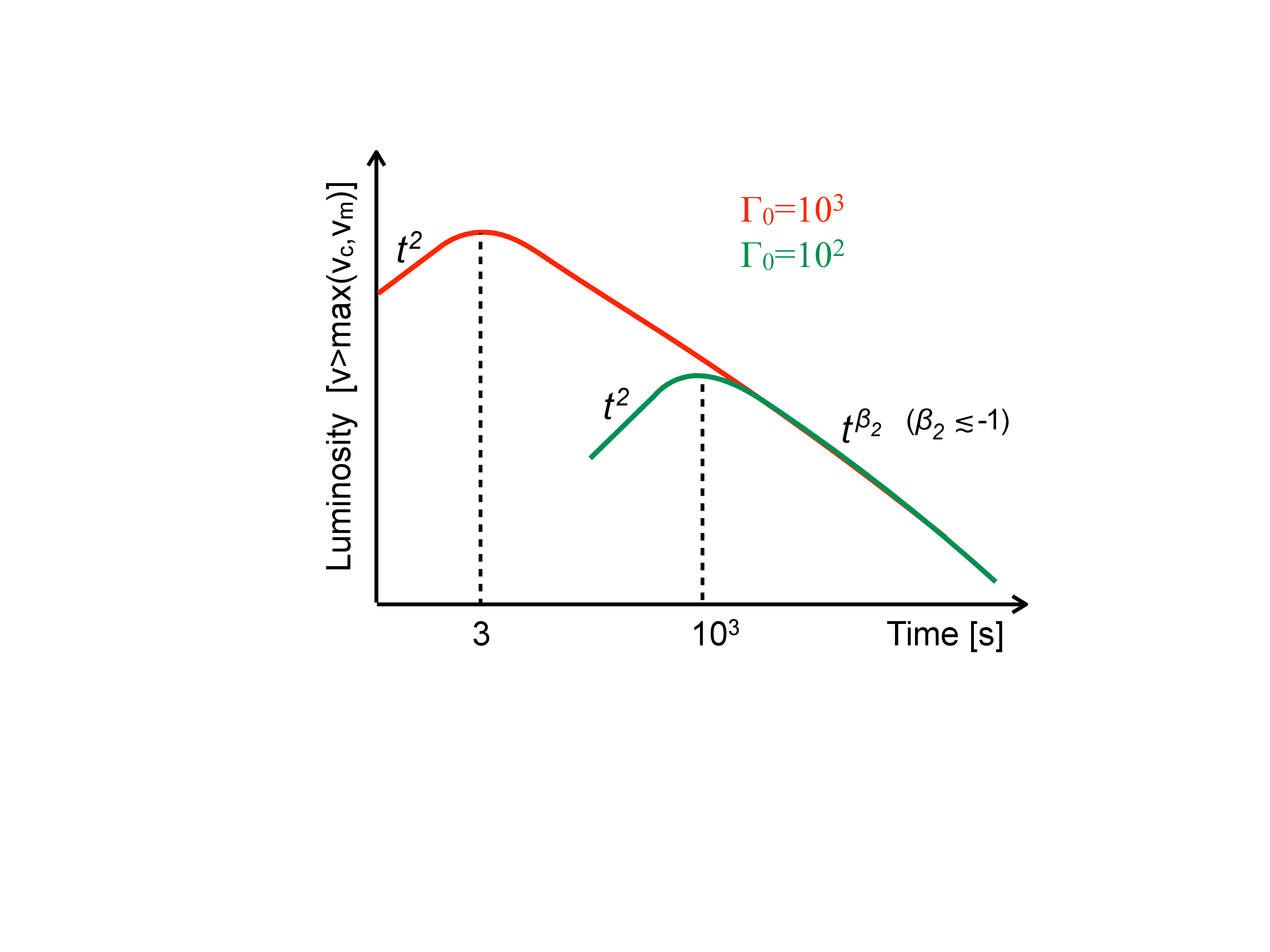}
\vskip -3.3truecm
\caption{Examples of  synchrotron afterglow lightcurves at a frequency $\nu>\max(\nu_{\rm c},\nu_{\rm m})$ for a constant density profile of the surrounding medium. The afterglow parameters are the same in both cases, except for the initial Lorentz factor \g0. At large $\Gamma_0\sim10^3$, the light curve peaks at early times (see equation~\ref{eq:peaktime_s=0}), while the peak is shifted at much later times when $\Gamma_0\sim10^2$. }
\label{fig:lc}
\end{figure}
While during the deceleration phase the value of \g0\ does not affect the flux (which is rather determined by the blast wave energy), \g0\ plays an important role during the coasting phase and in determining the deceleration time, i.e. the time of the transition from a constant to a decreasing Lorentz factor.
For small \g0, the deceleration occurs at late times and the peak flux is smaller.
To clarify this point, Fig.~\ref{fig:lc} illustrates the afterglow lightcurves of two GRBs that have the same parameters except for the initial Lorentz factors \g0. 
Even though they have the same energy \eiso\ (and hence same afterglow luminosity after deceleration), the chances to detect emission are very different in the two cases. Depending on the temporal window of observation as compared to the time of the peak, the afterglow of the low-\g0\ GRB might be completely missed. 
For a GRB observed within the first few hundred seconds, chances of detection are larger for high-\g0\ events. 
When observations extend to times longer than the peak time, where the luminosity is proportional to \eiso, the chances are dominated by the GRB energetics, and are larger for GRBs with a large \eiso.
From this example it is clear that three quantities play a fundamental role: the prompt energy \eiso, the observation time and \g0. When the first two quantities are known, a limit on \g0\ can be inferred from the non detection of the expected radiation.

\subsection{Synchrotron fluence at $\nu>\max(\nu_{\rm c},\nu_{\rm m})$}\label{sec:estimates}
Since the LAT is a photon-limited instrument, for a fixed spectral index $\alpha$ the detection capability is directly related to the fluence. 
We then estimate the synchrotron afterglow fluence \slat\ in the energy range $0.1-10\,$GeV (observer frame) under the assumption  $\max(h\nu_{\rm c},h\nu_{\rm m})<0.1\,$GeV. In this spectral range, the spectral slope $\alpha$ (in the notation $F_{\nu}\propto\nu^{\alpha}$) is $\alpha=-p/2$.
We model the external shock dynamics starting from the coasting phase, following \cite{nava13}, and the radiation output following \cite{sari98} and \cite{nappo14}.
The choice of computing the afterglow fluence in the range 0.1-10 GeV is motivated by the fact that available estimates of LAT flux upper limits have been computed in this energy range \citep{latcatalogul12}. Moreover, this is also the energy range chosen in the First {\it Fermi}-LAT GRB catalog \citep{latcatalog} to quote fluxes and fluences of LAT detected GRBs, that can be directly compared to the estimates provided in the following. We also note that, if extended up to higher energies ($>10\,$GeV), the estimates of the expected afterglow flux might significantly depend on the possible presence of a spectral cutoff, caused for example, by the maximum synchrotron energy. Limiting the estimates at energies smaller than $10\,$GeV reduces these uncertainties (see a discussion in section \ref{sec:suppression}).

We consider two different radial density profiles characterized as $n \propto r^{-s}$: a constant ($s=0$) and a decreasing density ($s=2$). 
While in both cases the afterglow flux after the deceleration time decreases with a temporal index $\beta_2=-(3p-2)/4$, before the deceleration the temporal indices are $\beta_1=2$ and $\beta_1=(2-p)/2$, for $s=0$ and $s=2$ respectively, where we used the notation $F(t)\propto t^\beta$. 
If observations start at $t_{\rm i}$ and end at $t_{\rm f}$ the fluence is:
\begin{eqnarray}
S^{\rm aft}_{[0.1-10]}=\int^{t_{\rm f}}_{t_{\rm i}}F^{\rm aft}_{[0.1-10]}\, dt~,
\label{eq:integral}
\end{eqnarray}
where the flux $F^{\rm aft}_{[0.1-10]}$ is given by:
\begin{eqnarray}
F^{\rm aft}_{[0.1-10]}\!=\!A
\left\{ \!
  \begin{array}{l}
 t^{\beta_1} \quad {\rm for}\ t\ll t_{\rm dec}\\
 \\
 t^{\beta_1}_{\rm dec} \left( \frac{t}{t_{\rm dec}}\right)^{-\frac{3p-2}{4}}\quad {\rm for}\ t\gg t_{\rm dec}\\
  \end{array} \right.
\end{eqnarray}
Here $t_{\rm dec}$ is the deceleration time in the observer frame. 
If observations are characterised by temporal gaps, the integration in eq.~\ref{eq:integral} should be performed separately in each time interval where observations are available. The total expected afterglow fluence will be the sum of the contributions from each time interval.

In what follows, we give analytic approximations of the numerical results for the computation of \slat\ (equation~\ref{eq:integral}), for different orders of the times $t_{\rm i}$, $t_{\rm f}$, and $t_{\rm dec}$.
We consider the general case $t_{\rm i}\ne0$, to account for cases where the GRB enters the LAT field-of-view (FoV) after the trigger time.

\subsubsection{Homogeneous medium: $n=constant$}
\begin{figure}
\vskip -0.25 truecm
\hskip -0.4truecm
\includegraphics[scale=0.51]{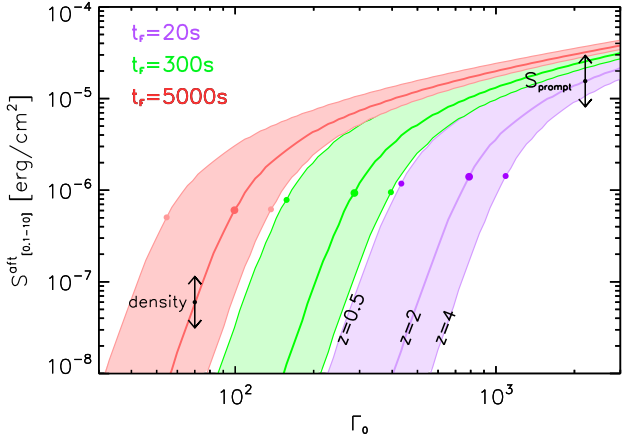}
\vskip -0.25 truecm
\caption{Synchrotron afterglow fluence integrated from $t_i=0$ to $t_{\rm f}$ in the range $0.1-10\,$GeV (observer frame). The three different stripes correspond to three different integration times $t_{\rm f}=20, 300, 5\times10^3\,$seconds (from right to left). For each stripe, the solid lines correspond to different redshifts: $z=0.5$ (upper boundary), $z=2$ (central thick line), and $z=4$ (lower boundary). The filled dots show the Lorentz factor for which the lightcurve peak time is equal to the integration time: $t_{\rm dec}=t_{\rm f}$. All the curves have been derived assuming \ee=0.1, \eb=0.01, \etag=0.2, $n$=1$\,$cm$^{-3}$, and \spr=$10^{-4}$erg/cm$^2$. Different values of \spr\ and $n_0$ significantly affect the curves, as indicated by the vertical arrows: in the first regime \slat\ depends almost linearly on the external density (see equation~\ref{eq:S_LAT_early_s=0}), while in the second regime the LAT fluence \slat\ depends linearly on the prompt fluence \spr\ (see equation~\ref{eq:S_LAT_late_s=0}). }
\label{fig:fluence_s=0}
\end{figure}
The transition from the coasting to the deceleration regime occurs around the deceleration time, which is also the time at which the lightcurve peaks:
\begin{equation}
t_{\rm dec}=3\,(1+z_2)^{2/3}\left[ \frac{S_{\rm \gamma,iso,-4}(1-\eta_{\gamma})\,d^2_{\rm L,2}}{\Gamma_{0,3}^8\,n_0\,\eta_{\gamma}} \right]^{1/3} \rm s~,
\label{eq:peaktime_s=0}
\end{equation}
where $S_{\rm \gamma,iso,-4}$ is the bolometric prompt fluence in units of $10^{-4}\,$erg/cm$^2$, and $n_0$ is the density in cm$^{-3}$. We use the notation $Q_x=Q/10^x$, except for the redshift (where $z_2$ means that the numerical factor has been estimated for a typical redshift $z=2$) and the luminosity distance $d_{\rm L,2}=d_{\rm L}/d_{\rm L,z=2}$.
We estimate the integral in equation~\ref{eq:integral} for three different cases: $t_{\rm dec} > t_{\rm f}$ (relevant for short observing times and/or for small values of \g0), $t_{\rm i}<t_{\rm dec}<t_{\rm f}$ (relevant for longer observing time and/or larger values of the Lorentz factor), and $t_{\rm dec}<t_{\rm i}$ (relevant when the GRB enters the FoV at late times, when the fireball is already decelerating).

\begin{itemize}
\item $t_{\rm dec} > t_{\rm f}$:
\begin{eqnarray}
\begin{array}{r}
S^{\rm aft}_{\rm [0.1-10]}=2.5\times 10^{-7}t^3_{\rm f,3}\Gamma_{0,2}^{(2p+4)}\epsilon_{\rm B,-2}^{\frac{p-2}{4}}n^{\frac{p+2}{4}}_0\epsilon_{\rm e,-1}^{p-1}\times\\
\\
\times(1+z_2)^{-\frac{p+2}{2}} d^{-2}_{\rm L,2} \left[1-\left(\frac{t_{\rm i}}{t_{\rm f}}\right)^3\right]\,\rm erg/cm^2~.
\end{array}
\label{eq:S_LAT_early_s=0}
\end{eqnarray}
In this first regime the dependence on \g0\ is very strong and there is no dependence on \spr. Moreover the fluence depends nearly linearly on $n_0$.\\

\item $t_{\rm i} < t_{\rm dec}<t_{\rm f}$:
\begin{eqnarray}
\begin{array}{c}
\label{eq:S_LAT_late_s=0}
S^{\rm aft}_{\rm [0.1-10]}=10^{-5} \rm erg/cm^2\, S_{\rm \gamma,iso,-4}\Gamma_{0,3}^{2(p-2)}\epsilon_{\rm B,-2}^{\frac{p-2}{4}}\,\times\\
\\
n^{\frac{p-2}{4}}_0\epsilon_{\rm e,-1}^{p-1}\frac{1-\eta_{\gamma}}{\eta_{\gamma}}(1+z_2)^{\frac{2-p}{2}}\times\\
\\
\left\{\left[1-\frac{4}{p+2}\left(\frac{t_{\rm f}}{t_{\rm dec}}\right)^{-\frac{3}{4}(p-2)}\right]- \frac{3(p-2)}{(p+2)}\left(\frac{t_{\rm i}}{t_{\rm f}}\right)^{-\frac{p-4}{2}}\right\} \,.
\end{array}
\end{eqnarray}
The dependences on $n_0$, \eb, and $z$ are very weak and can be neglected. 
Also, according to observations of GRBs with \tee, the term $\epsilon_{\rm e}^{p-1}\left[\frac{1-\eta}{\eta}\right]^{\frac{p+2}{4}}$ has a similar value for all GRBs \citep{nava14}. 
The main parameters determining the afterglow fluence are then \spr\ and (depending on the value of $p$) \g0.\\

\item $t_{\rm dec}<t_{\rm i}$:
\begin{eqnarray}
\begin{array}{c}
\label{eq:S_LAT_late_late_s=0}
S^{\rm aft}_{\rm [0.1-10]}=3\times10^{-5} S^{\frac{p+2}{4}}_{\rm \gamma,iso,-4}\epsilon_{\rm B,-2}^{\frac{p-2}{4}}\epsilon_{\rm e,-1}^{p-1}\left[\frac{1-\eta_{\gamma}}{\eta_{\gamma}}\right]^{\frac{p+2}{4}}\times\\
\\
\times\,d^{\frac{p-2}{2}}_{\rm L,2} t_{\rm i,3}^{-\frac{3(p-2)}{4}}\left[ 1-\left( \frac{t_{\rm f}}{t_{\rm i}} \right)^{-\frac{3(p-2)}{4}} \right]\rm erg/cm^2~.
\end{array}
\end{eqnarray}
In this last regime the synchrotron fluence is proportional to \spr\ but, contrary to the previous regime, it is independent of \g0.

\end{itemize}

The results are summarised in Fig.~\ref{fig:fluence_s=0}, that shows curves of \slat\ as a function of $\Gamma_0$. These have been derived for $t_{\rm i}=0$, but they hold as long as $t_{\rm i}<\min(t_{\rm dec},t_{\rm f})$, since for $n=const$ most of the emission is radiated at $t\gtrsim t_{\rm dec}$, and the initial integration time does not significantly affect the fluence estimates. 
Each shaded stripe corresponds to a different value of the final integration time $t_{\rm f}$ (from left to right: $t_{\rm f}=5\times10^3, 300, 20$ seconds). 
We chose $t_{\rm f}=5\times10^3$ as maximum value because this roughly corresponds to the maximum timescale over which observations can be performed without temporal gaps.
For each stripe, three different curves (corresponding to three different values of the redshift) are marked with a solid line: $z=0.5$ (upper boundary), $z=2$ (central thick line), and $z=4$ (lower boundary).
All curves have been derived for $S_{\rm \gamma,iso}=10^{-4}\,$erg cm$^{-2}$, \ee=0.1, \eb=0.01, $\eta_\gamma=0.2$, and $n=1\,$cm$^{-3}$.

Low values of $\Gamma_0$ correspond to late peak times. In this first regime, \slat\ strongly depends on \g0\ and on the redshift (see equation~\ref{eq:S_LAT_early_s=0}). Moreover, it depends nearly linearly on the density: the curves should be moved up/down for increasing/decreasing density, as indicated by the arrows. 
The prompt fluence plays no role in this regime. 

For increasing \g0\ the peak time decreases. For each curve, the \g0\ at which $t_{\rm dec}=t_{\rm f}$ is marked by a filled dot. At larger \g0
we switch to the regime $t_{\rm dec}<t_{\rm f}$. In this second regime the afterglow fluence depends very weakly on all the unknown parameters, except \spr. All the curves (for different $t_{\rm f}$ and redshifts) flatten (i.e. the dependence on \g0\ is weaker) and converge to a similar value, as predicted by equation~\ref{eq:S_LAT_late_s=0}. This value is  proportional to \spr: the curves should be moved up/down for increasing/decreasing prompt fluence, as indicated by the arrows. 

If a LAT observation results in a non-detection, and the upper limit on the LAT average flux is estimated on a time [$t_{\rm i},t_{\rm f}$], these plots and equations \ref{eq:S_LAT_early_s=0} to~\ref{eq:S_LAT_late_late_s=0} can be used to set an upper limit on \g0.
Under favourable observing conditions, the most stringent limits that LAT can place on the 0.1-10 GeV fluence are around a few$\times 10^{-7}\,$erg/cm$^2$ \citep{latcatalogul12,latcatalog}. Our calculations show that strong limits ($\lesssim 200$) on \g0\ can hence be placed only if the GRB is observed for at least several hundred seconds (green stripe in Fig.~\ref{fig:fluence_s=0}).

While the curves in Fig.~\ref{fig:fluence_s=0} have been derived under the assumptions that LAT observations start at the trigger time and that there are no temporal gaps in the observations, eqs.~\ref{eq:S_LAT_early_s=0} to~\ref{eq:S_LAT_late_late_s=0} can also be used in the more general case where $t_{\rm i}\ne0$ and/or in case of gaps during observations, for example caused by Earth occultation. In this latter case, the equations should be applied to each time interval where observations are performed, and the total fluence can then be estimated as the sum of contributions from each interval.

\subsubsection{Wind-shaped environment: $n\propto r^{-2}$}
We derive the synchrotron fluence at $\nu > \max(\nu_{\rm c}, \nu_{\rm m})$ for a density $n=3\times10^{35}A_{\star} r^{-2}$, where $A_{\star}$ is defined such that $A_\star=1$ corresponds to the case of a typical wind from a Wolf-Rayet star \citep{chevalier00}. 
The deceleration occurs around the time:
\begin{equation}
t_{\rm dec}=350\,\,\frac{S_{\rm \gamma,iso,-4}(1-\eta_{\gamma})\,d^2_{\rm L,2}}{\Gamma_{0,2}^4\,A_{\rm \star}\,\eta_{\gamma}}\,\, \rm s~.
\label{eq:peaktime_s=2}
\end{equation}

Also in this case, we consider all three possibilities for the order of $t_{\rm i}$, $t_{\rm f}$, and $t_{\rm dec}$.
Similar considerations to the case $s=0$ can be derived.\\

\begin{itemize}
\item $t_{\rm dec} > t_{\rm f}$:
\begin{eqnarray}
\begin{array}{r}
S^{\rm aft}_{\rm [0.1-10]}=3.7\times 10^{-6}t^{\frac{4-p}{2}}_{\rm f,3}\Gamma_{0,2}^{(p+2)}\epsilon_{\rm B,-2}^{(p-2)/4}A_{\rm \star}^{(p+2)/4}\times\\
\\
\times\epsilon_{\rm e,-1}^{p-1}d^{-2}_{\rm L,2}\left[ 1-\frac{t_{\rm i}}{t_{\rm f}}\right]^{\frac{4-p}{2}}\,\rm erg/cm^2~.
\end{array}
\label{eq:S_LAT_early_s=2}
\end{eqnarray}
\\

\item $t_{\rm i} < t_{\rm dec} < t_{\rm f}$:
\begin{eqnarray}
\begin{array}{r}
S^{\rm aft}_{\rm [0.1-10]}=10^{-5} \rm{erg/cm^2}\, S^{\frac{4-p}{2}}_{\rm \gamma,iso,-4}\Gamma_{0,2}^{3(p-2)}\times\\
\\
\times\,\epsilon_{\rm B,-2}^{\frac{p-2}{4}}A_{\rm \star}^{\frac{3(p-2)}{4}}\epsilon_{\rm e,-1}^{p-1}\left[\frac{1-\eta_{\gamma}}{\eta_{\gamma}}\right]^{\frac{4-p}{2}}d^{2-p}_{\rm L,2}\times\\
\\
\left\{\left[1-\frac{2(4-p)}{p+2}\left(\frac{t_{\rm f}}{t_{\rm dec}}\right)^{-\frac{3}{4}(p-2)}\right]- \frac{3(p-2)}{(p+2)}\left(\frac{t_{\rm i}}{t_{\rm f}}\right)^{-\frac{p-4}{2}}\right\}~,
\end{array}
\label{eq:S_LAT_late_s=2}
\end{eqnarray}
\\

\item $t_{\rm dec}<t_{\rm i}$:
\begin{eqnarray}
\begin{array}{r}
S^{\rm aft}_{\rm [0.1-10]}=3\times10^{-5} S^{\frac{p+2}{4}}_{\rm \gamma,iso,-4}\epsilon_{\rm B,-2}^{\frac{p-2}{4}}\epsilon_{\rm e,-1}^{p-1}\left[\frac{1-\eta_{\gamma}}{\eta_{\gamma}}\right]^{\frac{p+2}{4}}\times\\
\\
\times\,d^{\frac{p-2}{2}}_{\rm L,2} t_{\rm i,3}^{-\frac{3(p-2)}{4}}\left[ 1-\left( \frac{t_{\rm f}}{t_{\rm i}} \right)^{-\frac{3(p-2)}{4}} \right]\rm erg/cm^2~.
\end{array}
\label{eq:S_LAT_late_late_s=2}
\end{eqnarray}
\end{itemize}

\begin{figure}
\vskip -0.25 truecm
\hskip -0.35truecm
\includegraphics[scale=0.51]{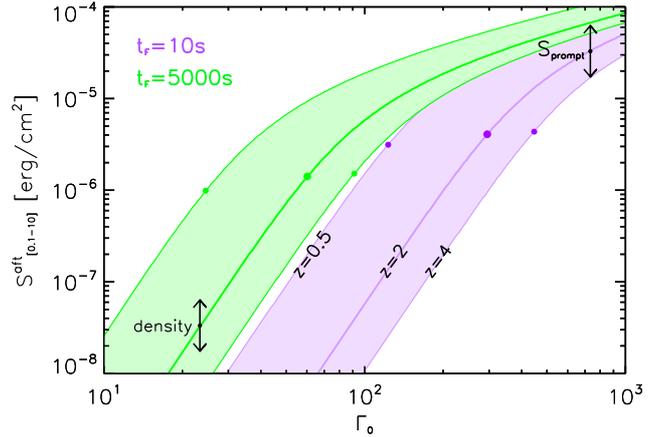}
\vskip -0.45 truecm
\caption{Same as in Fig.~\ref{fig:fluence_s=0} but for a wind circumburst density profile with $A_{\star}=1$ (see eqs.~\ref{eq:peaktime_s=2} to \ref{eq:S_LAT_late_s=2}). The two shaded stripes correspond to two different integration times: $t_{\rm f}=10, 5\times10^3\,$seconds (from right to left).}
\label{fig:fluence_s=2}
\end{figure}

The results are summarised in Fig.~\ref{fig:fluence_s=2}, that shows \slat\ as a function of \g0, for the case $t_{\rm i}=0$. Each shaded stripe corresponds to a different value of the final integration time $t_{\rm f}$. Since the dependence on $t_{\rm f}$ is weaker as compared to the case $n=n_0$, only two cases are shown: $t_{\rm f}=5\times10^3, 10$ seconds (from left to right). 
All curves have been derived for $S_{\rm \gamma,iso}=10^{-4}\,$erg cm$^{-2}$, \ee=0.1, \eb=0.01, $\eta_\gamma=0.2$, and $A_{\rm \star}=1$.
As in the constant density case, in the first regime ($t_{\rm f}<t_{\rm dec}$) the afterglow fluence depends on \g0\ and $z$, although the dependence on \g0\ is weaker (see equation~\ref{eq:S_LAT_early_s=2}). 
Moreover, it depends nearly linearly on the density: the curves should be moved up/down for increasing/decreasing density. 
The prompt fluence plays no role in this regime. For increasing \g0\ the deceleration time decreases and we switch to the regime $t_{\rm dec}<t_{\rm f}$.  For each curve, the \g0\ at which $t_{\rm dec}=t_{\rm f}$ is marked by a filled circle. In the second regime the fluence depends very weakly on all the unknown parameters, except \spr. All the curves converge to a similar value, as predicted by equation~\ref{eq:S_LAT_late_s=2}. This value is roughly proportional to \spr.

In the wind density scenario, LAT upper limits as deep as a few$\times10^{-7}\,$erg/cm$^{-2}$ lead to place stronger limits on \g0, as compared to the constant density case, even in the case of relatively short observation times $t_{\rm f}$.

\subsection{Caveats}\label{sec:suppression}
The estimates presented in the previous section neglect possible physical processes that might decrease the expected flux.
The high-energy synchrotron afterglow emission might indeed be affected by:

\begin{itemize}
\item Inverse Compton scattering: in this case the synchrotron luminosity at frequencies larger than $\max(\nu_{\rm c},\nu_{\rm m})$ is suppressed by a factor (1+$Y$), where $Y$ is the Compton parameter. This can be relevant for small values of \eb, a very uncertain parameter in GRB studies. However, at high-energies, the Compton scattering is in Klein-Nishina regime, and the relevance of inverse Compton effects is strongly reduced. \citet{beniamini15} have shown that $Y$ at $0.1-10\,$ GeV is of order unity, even for very small values ($<10^{-5}$) of \eb;

\item The maximum synchrotron photon energy: this is limited by the maximal energy up to which electrons can be shock-accelerated. 
The limit is estimated to be around $\Gamma\times70\,$MeV \citep{dejager92,piran10}. This means that the maximum photon energy is constant during the coasting phase and then it decreases. For $\Gamma<150$, this limit is then expected to produce a cutoff in the afterglow synchrotron spectrum around the energies relevant for this study. 
The extrapolation of the synchrotron spectrum with index $\alpha=-p/2$ up to $10\,$GeV, might then be incorrect. In this case the flux is smaller then what estimated before, especially at late times, when $\Gamma$ has significantly decreased. Since we will apply our estimates to early time observations ($t_{\rm f}=100\,$s, see section~\ref{sec:UL}) and since $\alpha<-2$, for the application presented here, this effect, if present, introduces a flux suppression at most of a factor of 2-3;

\item $\gamma$-$\gamma$ absorption: for LAT observations performed simultaneously to the prompt emission, it might be relevant to include $\gamma$-$\gamma$ absorption of GeV photons passing the shell of lower energy, prompt photons. Even though afterglow photons are produced at much larger radii as compared to prompt photons, \citet{zou11} have demonstrated that opacity might still arise and partially suppress the GeV flux.
\end{itemize}

All these processes, if relevant, lower the expected synchrotron fluence, as compared to estimates presented in the previous section. This would lead to higher upper limits on \g0\ (i.e., if the expected flux is smaller, non-detections are consistent with theoretical expectations also for higher \g0, leading to less stringent upper limits on \g0). This might be regarded as a weakness of the method. 
On the other hand, this can be used to check consistency by comparing
the upper limits derived with this method with lower limits and direct estimates derived with different methods. If the comparison does not outline any inconsistency, the assumption that the GeV afterglow flux is not strongly suppressed is well supported. On the other hand, an inconsistency between this and other methods would reveal the need for at least one of the mentioned processes to be at work.
As we will show later, inconsistencies are not found.

\section{Upper limits on $\Gamma_0$}\label{sec:UL}
As of January 2016 the GBM has detected prompt emission from almost $1800$\footnote{http://heasarc.gsfc.nasa.gov/W3Browse/fermi/fermigbrst.html} GRBs. Around 105 have been detected also by the LAT\footnote{http://fermi.gsfc.nasa.gov/ssc/observations/types/grbs/lat\_grbs/}, corresponding to around 13\% of the GRBs falling within the nominal LAT FoV, i.e. at an angle of $65^\circ$ from the LAT boresight (see also \citealt{vianello15}).
\begin{figure*}
\includegraphics[scale=0.68]{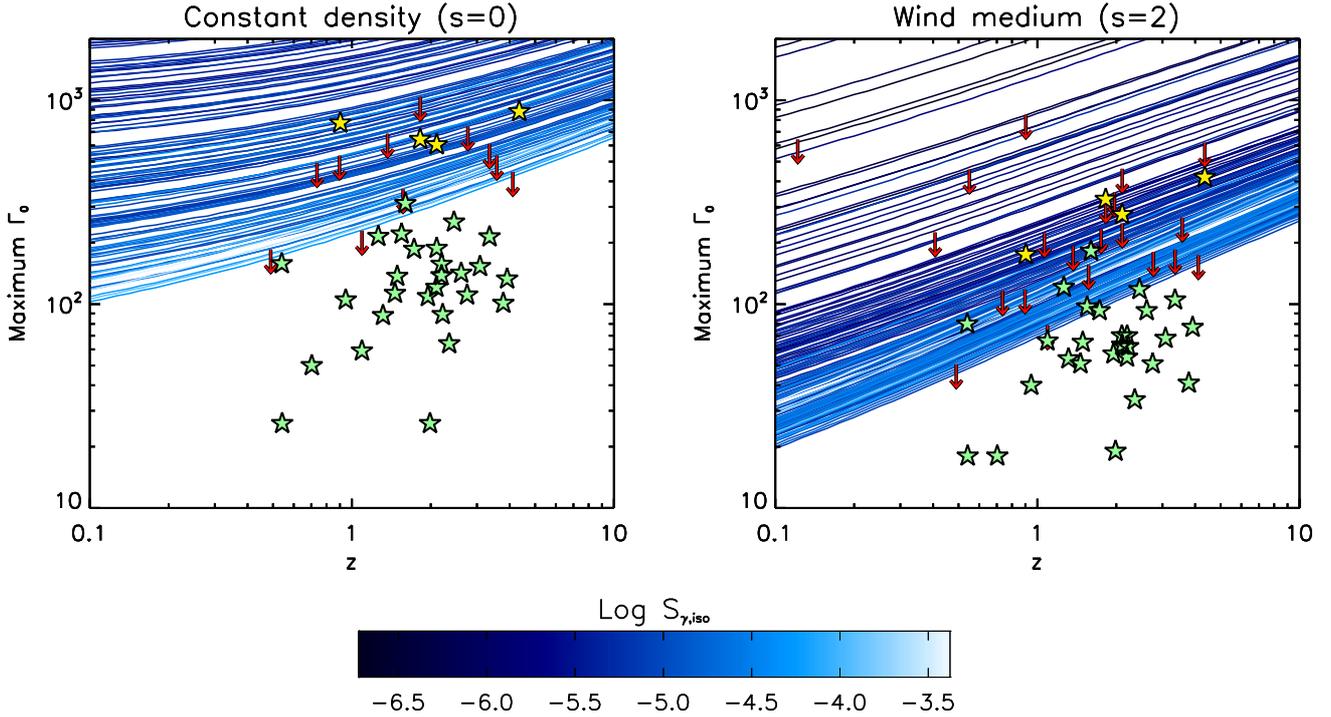} 
 \caption{Upper limits on $\Gamma_0$ for bursts with no LAT detection, as a function of the redshift, for a constant density medium (left panel) and a wind shaped medium (right panel). Only upper limits smaller than $\Gamma_0=2000$ are shown. Different colours of the curves refer to different values of the prompt GBM fluence: lighter colours are used for brighter bursts (see the color bar). Red arrows: upper limits for GRBs with measured redshift. Star symbols: GRBs for which $\Gamma_0$ has been estimated from the peak of the early optical lightcurve (green stars) and GeV light curve (yellow stars), taken from \citet{ghirlanda12}. }
 \label{fig:G0limits}
\end{figure*}
 \cite{latcatalogul12} have considered all GRBs with no evidence of emission above 100 MeV, that fell within the LAT FoV during the first 2.5 years (288 events).  
The upper limits on the average flux in the range 0.1-10~GeV have been estimated on three different integration times: during the prompt emission, and for fixed 30 s and 100 s integration times, starting from the trigger time (i.e. $t_{\rm i}=0$). 
We consider here the upper limits estimated for $t_{\rm f}=100\,$s. 
For each burst in this sample, we have computed the prompt fluence \spr\ in the energy range $1-10^4\,$keV using the best fit model reported in the {\it Fermi} GBM burst online catalog\footnote{http://heasarc.gsfc.nasa.gov/W3Browse/fermi/fermigbrst.html} (\citealt{bhat16}).
The fit models used in the catalog include a simple power-law (PL), a power-law with an exponential cutoff (CPL), a smoothly broken PL (SBPL), and the so-called Band function. We have considered only those GRBs for which the best fit model is a peaked (in $\nu F_\nu$) function (i.e. either the CPL, SBPL, or Band models), otherwise a model extrapolation down to $1\,$keV and up to $10\,$MeV would be unsafe.
The final sample includes 190 GRBs.
In this sample we find that the limits on the LAT fluence  $S^{UL}_{\rm [0.1-10]}$ in the first 100 seconds range from $5\times10^{-7}\,$erg/cm$^2$ to $8\times10^{-5}\,$erg/cm$^2$.

For this sample, $t_{\rm i}=0$, and $t_{\rm f}$, \spr, and $S^{UL}_{\rm [0.1-10]}$ are known. Imposing \slat$<S^{UL}_{\rm [0.1-10]}$, equations~\ref{eq:S_LAT_early_s=0} and \ref{eq:S_LAT_late_s=0} (or \ref{eq:S_LAT_early_s=2} and \ref{eq:S_LAT_late_s=2} for the wind case) can then be inverted to find the upper limit on \g0. 
We assume \ee=0.1, \etag=0.2, $n_0=1$ (or $A_\star=1$ for the wind density case), \eb=0.01 and $p=2.2$ (but \eb\ and $p$ do not affect the estimates, and the density is important only at small values of \g0). 
Since the redshift is known only for a small fraction of the sample, we derive the upper limits on \g0\ as a function of $z$. 
The results are shown in Fig.~\ref{fig:G0limits} both for a constant density medium (left panel) and a wind medium (right panel), for $z$ in the range 0.1-10 (blue and light-blue curves). Red arrows mark those bursts for which the redshift is known.
Only cases resulting in upper limits smaller than 2000 are shown.
To emphasise the role of the prompt fluence, we use different colours for different values of \spr: brighter (in the GBM range) bursts are marked with lighter colours. It is evident that stringent limits on \g0\ can be derived only for the brightest GRBs.
For a typical redshift $z\sim2$, the limits on $\Gamma_0$ lie above $200$ and in the range 100-400 for a constant and wind-like medium, respectively.

These limits can be compared with limits and direct estimates available in the literature and computed with different methods. 
\cite{latcatalogul12} derived upper limits for 6 bright GRBs for which a high-energy cutoff in the prompt spectrum at energies $<100\,$MeV is implied by the LAT non detection.
Their upper limits on \g0\ as a function of $z$ are shown in their figure 11. The curves are similar to those derived here, with limiting values around $\sim$150 at $z=0.5$ and $\sim$500 at $z=5$.
Upper limits on \g0\ have been computed also from early time X-ray observations, resulting in maximum values around several hundreds, by \cite{zou10}. They have also shown that when these are combined with lower limits required to avoid the compactness problem, values of \g0\ are in the range $10^2-10^3$.

Concerning direct estimates (rather then limits) of \g0, a spectral break in the prompt component has been observed only in a few cases \citep{090926ALAT,tang15}. 
Most of the available estimates of the value of \g0\ have been inferred from the detection of an early peak in the afterglow lightcurve.
\cite{ghirlanda12} collected all GRBs with known redshift and with an early peak in the optical light curve, and inferred \g0\ under the assumption that the peak corresponds to the blast wave deceleration time.
The \g0\ values have been derived both for a constant and wind-like medium, and are shown in Fig.~\ref{fig:G0limits} as star symbols (the green colour refers to optical lightcurves, while the yellow colour refers to a similar analysis applied to GeV lightcurves of LAT GRBs with temporally extended GeV emission).

The most stringent limits derived in this work lie above most of the values inferred from GRBs with an optical peak.
This implies that the non-detection of synchrotron afterglow radiation is consistent with the simplest model, and there is no evidence that mechanisms producing a suppression of the GeV flux (see section~\ref{sec:suppression}) are at work. The possibility to test the relevance of these processes is however limited by the instrument sensitivity. We can conclude that present instrument capabilities are not pointing to the need for a relevant suppression of the high-energy afterglow synchrotron flux. 

On the other hand, the upper limits lie not far from (and sometimes below) the estimated values of \g0. This suggests that the LAT should be able to detect the synchrotron afterglow component for those GRBs with the largest bulk Lorentz factors and largest energetics.
A fraction of the LAT detected GRBs are indeed characterised by the presence of an emission above 100$\,$MeV lasting much longer than the prompt radiation, whose flux decays in time as a power-law \citep{latcatalog}. These are the brightest GBM GRBs, and a large Lorentz factor $\Gamma_0>500$ has been inferred for them. An association with synchrotron afterglow radiation has been claimed to be consistent also with their spectral and temporal properties \citep{kumar09,kumar10,ghisellini10,ghirlanda10,depasquale10,lemoine13,nava14,beniamini15}, although photons with particularly large energies ($>10$\,GeV) detected at late times ($>10^2\,$s) are in excess of the synchrotron limit and require a different explanation \citep{piran10,wang13,130427LAT}. 

Finally, we comment on the dependence of these results on the unknown parameters \ee\ and \etag, with reference to a homogeneous density medium (similar considerations hold also for a wind-shaped density medium). In the first regime, where observations stop before the lightcurve reaches the peak (equation~\ref{eq:S_LAT_early_s=0}), our estimates of \g0\ do not depend on \etag, and they depend very weakly on \ee\ (\g0$\propto\epsilon_{\rm e}^{0.1}$). In the second regime (equation~\ref{eq:S_LAT_late_s=0}), the Lorentz factor appears also in the definition of $t_{\rm dec}$, and it is then less obvious to understand how different assumptions on \etag\ and \ee\ affect the results. From numerical estimates, we find that if the value of $\epsilon_{\rm e}(1-\eta_\gamma)/\eta_\gamma$ increases by a factor of 5 compared to our fiducial value of 0.4, the upper limits on \g0\ are smaller, by a factor of 1.5. They lie closer to the direct values estimated from the peak of optical lightcurves (green star symbols in Fig.~\ref{fig:G0limits}). Conversely, if the value of $\epsilon_{\rm e}(1-\eta_\gamma)/\eta_\gamma$ is decreased by a factor of 10, the upper limits increase by a factor of $\lesssim3$, and the most stringent values are now at the level of the direct estimates derived from GeV lightcurves of GRBs with LAT temporally extended emission (yellow star symbols in Fig.~\ref{fig:G0limits}).

\section{Implications for GRBs with detected GeV temporally extended emission}
\label{sec:detected}
\begin{figure*}
\hskip -0.4truecm \includegraphics[scale=0.83]{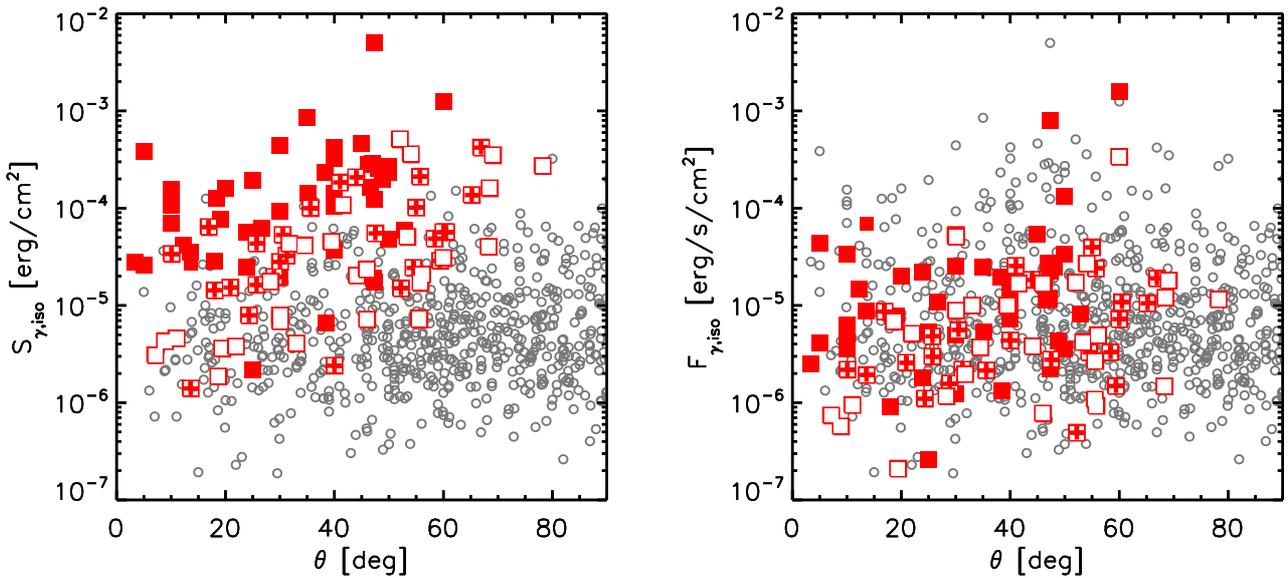} 
 \caption{Prompt fluence (left panel) and peak flux (right panel) in the energy range 1-$10^4\,$keV vs. the angle $\theta$ to the LAT boresight. Grey dots represent {\it Fermi} GRBs detected only by the GBM. Square symbols represent GRBs detected also by the LAT: filled symbols refer to those with temporally extended emission, empty symbols refer to those with no evidence for extended emission, empty symbols with a cross inside refer to cases for which the classification is uncertain.}
 \label{fig:prompt-angle}
\end{figure*}
In order to detect synchrotron afterglow radiation with the LAT, \slat\ must be larger than the instrument threshold:
\begin{equation}
S^{\rm aft}_{\rm [0.1-10]}>S_{\rm th}[\rm bkg,\theta, t_{\rm f}-t_{\rm i}, \alpha].
\label{eq:condition}
\end{equation}
This threshold does not have the same value for all GRBs, because it strongly depends on the specific observing conditions. It is then impossible to identify a unique condition that all GRBs must satisfy in order to have a detectable GeV afterglow radiation.
More precisely, the minimum value of the fluence $S_{\rm th}$ required for detection will in general depend on the level of background, the angle $\theta$ between the burst location and the LAT boresight (which might also change during observations), 
how long the GRB is inside the LAT FoV ($\Delta t=t_{\rm f}-t_{\rm i}$), and the spectral index $\alpha$. 
The level of background depends on contamination from earth-albedo events CR-background, on the geomagnetic latitude, and on the location of the Earth limb.
$S_{\rm th}$ can then considerably vary from burst to burst, and two events with similar intrinsic properties and located at similar distances can result in a detection or non-detection due to different observing conditions.

As discussed in section~\ref{sec:estimates}, also the theoretical estimate of \slat\ cannot be fully determined, because it depends on a few unknown parameters, such as \g0, $z$ (which is not measured in most cases) and possibly $n$ (depending on the interval time $t_{\rm i}-t_{\rm f}$ during which the event is observed).
However, for a typical $t_{\rm f}$ (of at least few hundred seconds) and for reasonably large Lorentz factors ($\Gamma_0>100$), the main parameter determining the afterglow fluence in the LAT range is the prompt fluence \spr\ (see Figs.~\ref{fig:fluence_s=0} and \ref{fig:fluence_s=2}, and equations~\ref{eq:S_LAT_late_s=0} and \ref{eq:S_LAT_late_s=2}), if \ee\ and \etag\ do not vary significantly. 
The condition for having a detectable afterglow fluence can then be roughly translated into a condition on the prompt fluence. 
Keeping in mind that this is true only in the regime $t_{\rm f}>t_{\rm dec}>t_{\rm i}$ and that also \g0\ plays a role in determining the afterglow fluence, the prediction is that, when the emission detected by LAT is indeed afterglow radiation, these events should also be the ones with the largest prompt fluences.
A correlation between the prompt sub-MeV fluence and the GeV fluence arises also if both emissions are related to the prompt component, but in this case it is not trivial to explain why the GeV radiation extends in time significantly beyond the prompt phase.

Following these considerations we collect all GRBs detected by {\it Fermi} up to January 2016 and plot their distribution in the plane \spr-$\theta$ (Fig.~\ref{fig:prompt-angle}, left panel), to verify if LAT detected GRBs with temporally extended emission show indeed a tendency to have larger prompt fluences.
For each burst in this sample, the prompt fluence \spr\ has been estimated in the energy range $1-10^4\,$keV using the best fit model reported in the {\it Fermi} GBM burst online catalog\footnote{http://heasarc.gsfc.nasa.gov/W3Browse/fermi/fermigbrst.html}.
Grey empty circles are GRBs detected by the GBM but with no emission detected by the LAT. 
GRBs detected also by the LAT\footnote{http://fermi.gsfc.nasa.gov/ssc/observations/types/grbs/lat\_grbs/} are instead marked with a square symbol.
Note that, when GeV radiation is detected, there is a possibility that this radiation is not synchrotron afterglow emission, i.e. there are cases where the afterglow fluence is too faint, and photons of a different origin (for example, the high-energy extension of the prompt spectrum) are responsible for the LAT detection.
To account for this possible contamination, we classify LAT GRBs according to the duration of the LAT emission as compared to the duration of the prompt detected by the GBM. 
According to information derived either from the GRB LAT catalog \citep{latcatalog}, the GCN archive, or literature, we divide the sample into three categories: i) GRBs with temporally extended emission (filled squares), ii) GRBs with no LAT emission after the end of the prompt emission (empty squares) and iii) GRBs for which the classification is uncertain, since a few photons have been detected by the LAT after the end of the prompt emission, but on timescales comparable to the prompt duration (squares with a plus symbol inside).

LAT GRBs with \tee\ and non-LAT GRBs clearly populate two different regions of the plane, with LAT GRBs to clustered in the high-\spr/low-$\theta$ region (Fig.~\ref{fig:prompt-angle}, left panel). 
We check if this tendency is present also when the prompt fluence is replaced with the prompt peak flux $F_{\rm\gamma,iso}$. The right panel in Fig.~\ref{fig:prompt-angle} shows, for the same sample, the prompt peak flux as a function of $\theta$. 
GRBs detected by the LAT now span almost all the range of peak fluxes, and no clear separation is present between LAT and GBM-only GRBs, indicating that the prompt peak flux does not influence the possibility of having a bright long-lasting high-energy component.
The separation between LAT and non-LAT bursts is instead evident in terms of prompt fluence, consistently with the afterglow model.

\section{Discussion and Conclusions}\label{sec:conclusions}
The luminosity of the early afterglow emission strongly depends on the value of the initial Lorentz factor \g0. This parameter indeed affects the expected emission in two ways: (i) it is the main parameter determining the deceleration time, i.e. the transition between the coasting phase (where the Lorentz factor is constant) and the deceleration phase, and (ii) it is the main parameter determining the luminosity of the radiation during the initial coasting phase. 
Large values of \g0\ imply a short deceleration time and a large peak flux. Afterglows of high-\g0\ GRBs are then easier to detect (Fig.~\ref{fig:lc}).
Early time flux upper limits can then be translated into upper limits on the afterglow luminosity, and in turn on the value of \g0.

In principle this method can be applied to optical and X-ray observations. The optical band, however, likely lies below the cooling frequency, where the flux depends on very uncertain parameters, especially \eb. 
Recent afterglow modelings on different samples selected in different energy bands (radio, optical, X-ray, and GeV) have showed that \eb\ probably spans a large range of values, covering at least 4-5 orders of magnitude \citep{barniolduran14,santana14,zhang15,lemoine13,beniamini15}, making the predictions of the optical flux very uncertain.
The X-ray band instead, lies most likely above the cooling frequency, but, for small values of \eb, this part of the synchrotron spectrum is strongly affected by inverse Compton scattering \citep{beniamini15,beniamini16}. Again, the very uncertain value of \eb\ would reflect on a large uncertainty on the expected X-ray flux, and then in not very robust limits on \g0. Higher ($\sim$GeV) energies are less affected by these issues: first, we can safely assume that the LAT energy range is above the cooling frequency, and second, the Klein-Nishina cross section strongly limits the effects of the inverse Compton scattering on this part of the synchrotron spectrum.

We have modeled $\sim$GeV synchrotron afterglow emission during the coasting and deceleration phases, and compared model expectations with LAT observations. Since the LAT is a photon limited instrument, for a fixed photon index the relevant quantity for the detection is the fluence. We have presented equations to estimate the synchrotron afterglow fluence in the range 0.1-10~GeV (observer frame) as a function of all afterglow parameters, prompt fluence, redshift, and initial ($t_{\rm i}$) and final ($t_{\rm f}$) observation times (see equations~\ref{eq:peaktime_s=0} to \ref{eq:S_LAT_late_late_s=0} for a homogenous density medium, and equations~\ref{eq:peaktime_s=2} to \ref{eq:S_LAT_late_late_s=2} for a wind-like density medium). 
For the case $t_{\rm i}=0$ (i.e., for GRBs that are inside the LAT FoV at the trigger time) the results are summarized in Fig.~\ref{fig:fluence_s=0} and \ref{fig:fluence_s=2} (for a constant and a wind-like density profile, respectively).
The fluence is shown as a function of \g0\ for different observing times $t_{\rm f}$ and for fixed \ee=0.1, \eb=0.01 and $p=2.2$ (the last two parameters however play a very little role in modifying the estimates), while the dependence on $n$, $z$, and \spr\ are shown in the figures.
These curves and the equations provided in section \ref{sec:estimates} can be used to set a limit on \g0, if the upper limit on the LAT average flux from $t_{\rm i}$ to $t_{\rm f}$ is known. 

We have applied these equations to a sample of 190 GRBs with no evidence for GeV emission \citep{latcatalogul12}.
We have used the upper limits on the average LAT flux (estimated in the first 100 seconds after the GRB trigger) to place upper limits on \g0\ (Fig.~\ref{fig:G0limits}).
For a typical redshift $z=2$, the inferred values are above 200 for a homogeneous medium, and in the range 100-400 for a wind-like density medium.
These values are consistent with estimates (and lower limits) available in literature and inferred with different methods.

These estimates rely on the assumption that processes such as the existence of a limit on the maximal synchrotron photon energy, $\gamma-\gamma$ absorption with lower energy, prompt photons, and inverse Compton scattering, do not significant lower the expected high-energy synchrotron flux (see section \ref{sec:suppression} for a discussion).
The lack of conflict between our results inferred from high-energy observations and estimates inferred (with other methods) from observations at lower frequencies implies there is no need to invoke a suppression of the high-energy afterglow flux. 
An improved instrument sensitivity is required to probe the presence and relevance of the mentioned processes.

On the other hand, the fact that most of the inferred upper limits lie very close to \g0\ values (and lower limits) estimated with different methods (see Fig.~\ref{fig:G0limits}, star symbols) implies that the synchrotron afterglow radiation from GRBs with the highest \g0\ should be bright enough to be detected by the LAT. 
For very large \g0, the deceleration time is small, and the afterglow luminosity is a good proxy for the blastwave energy. In turn, for a fixed value of the prompt efficiency \etag, the blastwave energy is a proxy for the energy radiated during the prompt, \eiso, implying that the GeV synchrotron radiation should be detectable for the GRBs with the highest prompt fluences.
This scenario is consistent with the detection, in a considerable fraction of the LAT GRBs, of a slowly fading GeV radiation on timescales much longer than the prompt emission, whose luminosity is tightly correlated with the prompt energy \citep{nava14}.
 
The synchrotron afterglow scenario is then consistent not only with detections, but also with non-detections of GRBs by the LAT, and offers a method to place upper limits on \g0. This method, in combination with other estimates (mostly lower limits), provides a tool to restrict the acceptable range of values for the still uncertain parameter \g0.

\section*{Acknowledgements}
LN was partially supported by a Marie Curie Intra-European Fellowship of the European Community's 7th Framework Programme (PIEF-GA-2013-627715).
LN and TP were supported by the ISF-CHE I-Core center for Excellence for research in Astrophysics (1829/12), 
and by the China-NSF-Israel-ISF grant.

\bibliography{biblio.bib}

\label{lastpage}

\end{document}